\documentclass[letterpaper,prb,twocolumn,amsmath,amssymb,floatfix]{revtex4-1}
\usepackage[T1]{fontenc}
\usepackage[latin9]{inputenc}
\usepackage{lmodern}
\usepackage{microtype,bbm}
\usepackage{graphicx}
\usepackage[usenames,dvipsnames]{xcolor}

\graphicspath{{../figs/}}

\usepackage{hyperref}
\hypersetup{
	colorlinks,
	allcolors=NavyBlue,
	linktoc=all
}

\usepackage[capitalize]{cleveref}

\renewcommand{\Re}{\mathrm{Re}}

\newcommand{\D}{\mathcal D}
\newcommand{\I}{\mathcal I}

\newcommand{\R}{\mathbb R}
\renewcommand*{\P}{\mathrm P}
 
\newcommand*\dagg{^{\dagger}}
\newcommand*\eps{\varepsilon}

\newcommand*\dens{{\rho}}
\newcommand*\mat[1]{\begin{pmatrix}#1\end{pmatrix}}

\DeclareMathOperator{\tr}{tr}
\let\Re\relax

\DeclareMathOperator{\Re}{Re}

\def\XXint#1#2#3{{\setbox0=\hbox{$#1{#2#3}{\int}$}
\vcenter{\hbox{$#2#3$}}\kern-.5\wd0}}

\begin{document}

\title{Current rectification in double quantum dot through fermionic reservoir engineering}
\author{Daniel Malz}
\author{Andreas Nunnenkamp}
\affiliation{Cavendish Laboratory, University of Cambridge, Cambridge CB3 0HE, United Kingdom}
\date{\today}

\begin{abstract}
  	Reservoir engineering is a powerful tool for the robust generation of quantum states or transport properties.
  	Using both a weak-coupling quantum master equation and the exact solution, we show that directional transport of electrons through a double quantum dot can be achieved through an appropriately designed electronic environment.
  	Directionality is attained through the interference of coherent and dissipative coupling.
  	The relative phase is tuned with an external magnetic field, such that directionality can be reversed, as well as turned on and off dynamically. 
	Our work introduces fermionic reservoir engineering, paving the way to a new class of nanoelectronic devices.
\end{abstract}

\maketitle

\section{Introduction}
Transport through nanoelectronic structures has been a thriving research field for many years, with quantum dots (QDs) being a prime example \cite{Wiel2002}.
Goals of this effort include high precision currents from single-electron pumps~\cite{Martinis1994,Chorley2012,Connolly2013,Pekola2013,VanZanten2016} and quantum devices encoding information with single electrons~\cite{Loss1998,SingleChargeTunneling,Tougaw1994}.
One important aspect of transport is current rectification.
It can be achieved through the Pauli spin blockade in double quantum dots (DQDs)~\cite{Ono2002,Fransson2006,Muralidharan2007,Buitelaar2008}
or through Coulomb blockade in triple quantum dots~\cite{Stopa2002,Vidan2004}.
In both cases, rectification is a result of many-body effects with an electron trapped permanently in one of the QDs.

Reservoir engineering promises robust generation of quantum states through designed environments~\cite{Poyatos1996}.
It has been applied to trapped atoms~\cite{Krauter2011}, trapped ions~\cite{Barreiro2011,Lin2013,Kienzler2015}, circuit quantum electrodynamics~\cite{Murch2012,Shankar2013,Leghtas2015} and cavity optomechanics~\cite{Kronwald2013,Woolley2014,Wollman2015,Pirkkalainen2015,Lecocq2015,Toth2017}. Recently, it has been exploited for promising magnetic-field-free directional devices for photons~\cite{Ranzani2014,Metelmann2015,Fang2017,Bernier2017,Peterson2017,Barzanjeh2017}.
Surprisingly, fermionic reservoir engineering is virtually unexplored, except for situations where the system couples to spin~\cite{Schuetz2014} or bosonic degrees of freedom~\cite{Grimsmo2016} of the reservoir.

In this Article, we present a novel mechanism for rectification in a DQD that works on the single-particle level and relies on dissipation in a reservoir shared between both dots.
In contrast to Refs.~\cite{Schuetz2014,Grimsmo2016}, the engineered reservoir exchanges fermions with the system. 
The mechanism is based on a directional interaction that arises due to interference of \emph{coherent} (from a Hamiltonian)
and \emph{dissipative} coupling (from a shared reservoir), independently of particle statistics~\cite{Metelmann2015}.
The relative phase of coherent and dissipative coupling is controlled by an externally applied magnetic field and can be tuned to yield forward directionality, backward directionality or reciprocal transport.
It is therefore a form of passive coherent control, in contrast to active feedback control~\cite{Brandes2010,Emary2016}, with potentially interesting consequences for quantum thermodynamics~\cite{Pekola2015}.

We unearth the directionality mechanism using a simple weak-coupling quantum master equation (QME) and
corroborate our analysis with the exact solution obtained from the Laplace transform of the equations of motion, which shows that the current-voltage characteristics are smoothed out over the width of the energy levels.
Finally, we discuss experimental implementation, and the impact of other physical effects on directionality, including non-Markovianity of the reservoir.
Our work introduces fermionic reservoir engineering, paving the way to a new class of nanoelectronic devices, with applications in electronic quantum information technology and precision current generation.

\section{Model}
\begin{figure}[t]
	\centering
	\includegraphics[width=\linewidth]{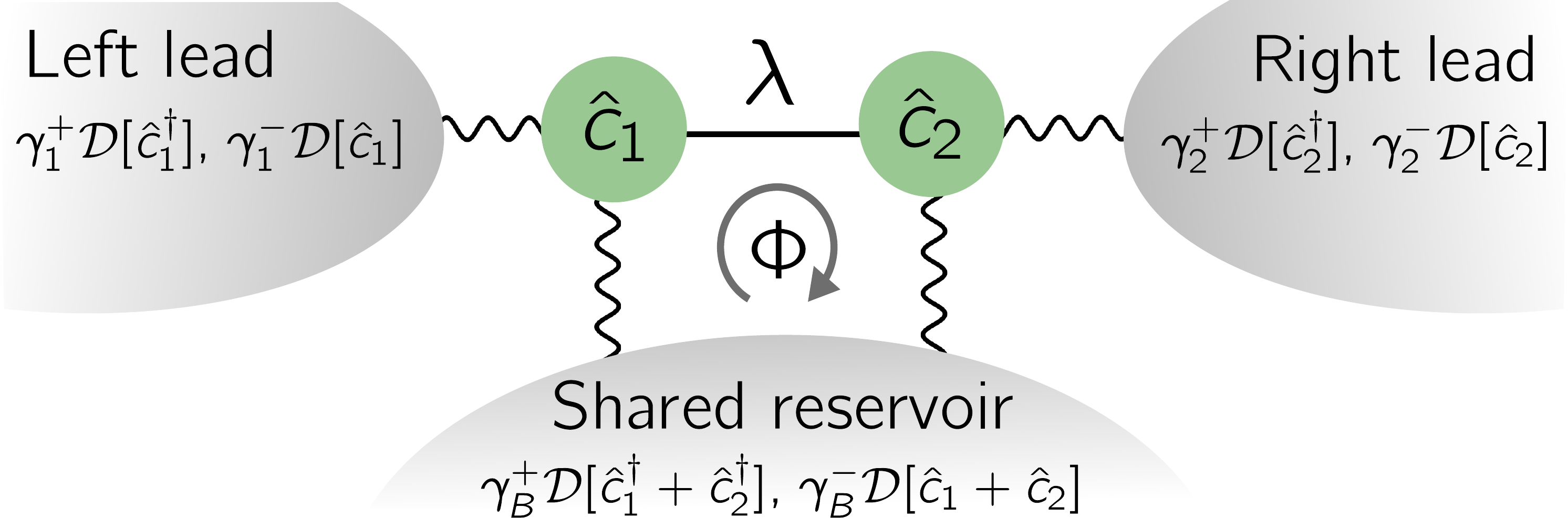}
	\caption{Schematic showing a double quantum dot (DQD) in contact with three reservoirs. We consider a single energy level in each dot, with annihilation operator $\hat c_1,\hat c_2$. Electrons can tunnel between the two sites with complex amplitude $\lambda$.
	Each dot is tunnel-coupled to a reservoir (denoted left and right lead, playing the role of source and drain) whose chemical potential can be controlled by externally applied voltages.
	The crucial feature of our proposal is that both sites are additionally tunnel-coupled to a shared reservoir that induces non-local electron loss.
	}
	\label{fig:model}
\end{figure}

We consider a serial DQD in a magnetic field, where each site is tunnel-coupled to a lead, and both are connected to a shared electronic reservoir (see \cref{fig:model}).
We assume that the energy level spacing in each dot is large compared to other parameters in the problem and that the chemical potentials are sufficiently low such that we only need to consider one level per dot.
If the applied magnetic field induces a large energy splitting between the spin states, such that only one spin state is relevant, we can drop the spin index.
Under these assumptions, the Hamiltonian of the system is ($\hbar=1$)
\begin{subequations}
	\begin{align}
		\hat H&=\hat H_{\text{sys}}+\hat H_{\text{res}}+\hat H_{\text{sys-res}},\\
		\hat H_{\text{sys}}&=\sum_{i=1}^2\eps_i \hat n_i+\lambda \hat c_1\dagg \hat c_2+\lambda^*\hat c_2\dagg\hat c_1,\\
		\hat H_{\text{res}}&=\sum_{\alpha=1,2,B}\sum_k \eps_{k,\alpha}\hat b_{k,\alpha}\dagg \hat b_{k,\alpha},\\
		\hat H_{\text{sys-res}}&=-\sum_k\sum_{i=1}^2\hat c_i\dagg \left( G_{k,i}\hat b_{k,B}+J_{k,i}\hat b_{k,i} \right)+\text{H.c.}
	\end{align}
	\label{eq:Hamiltonian}
\end{subequations}
Here, $\hat n_i=\hat c_i\dagg\hat c_i$ is the fermionic number operator for site $i$, $\lambda$ the complex tunneling amplitude between the dots, $\hat b_{k,\alpha}$ are the annihilation operators for fermions in the reservoirs, and $G_{k,i},J_{k,i}$ are real couplings of the sites to the reservoir modes.

In presence of a magnetic field, electrons moving in a closed loop pick up a phase proportional to the flux through the loop. In our system, the only closed loop is formed by the two dots with the shared reservoir (cf.~\cref{fig:model}).
In \cref{eq:Hamiltonian} we have chosen a gauge in which the resulting Peierls phase $\Phi$ is associated with the inter-dot coupling $\lambda=|\lambda|\exp(i\Phi)$.
This phase is the crucial ingredient to obtain destructive interference between coherent and dissipative interaction.
While time-reversal symmetry is broken by dissipation, the applied magnetic field breaks the symmetry under exchange of 1 and 2.

\begin{figure}[t]
	\centering
	\includegraphics[width=.75\linewidth]{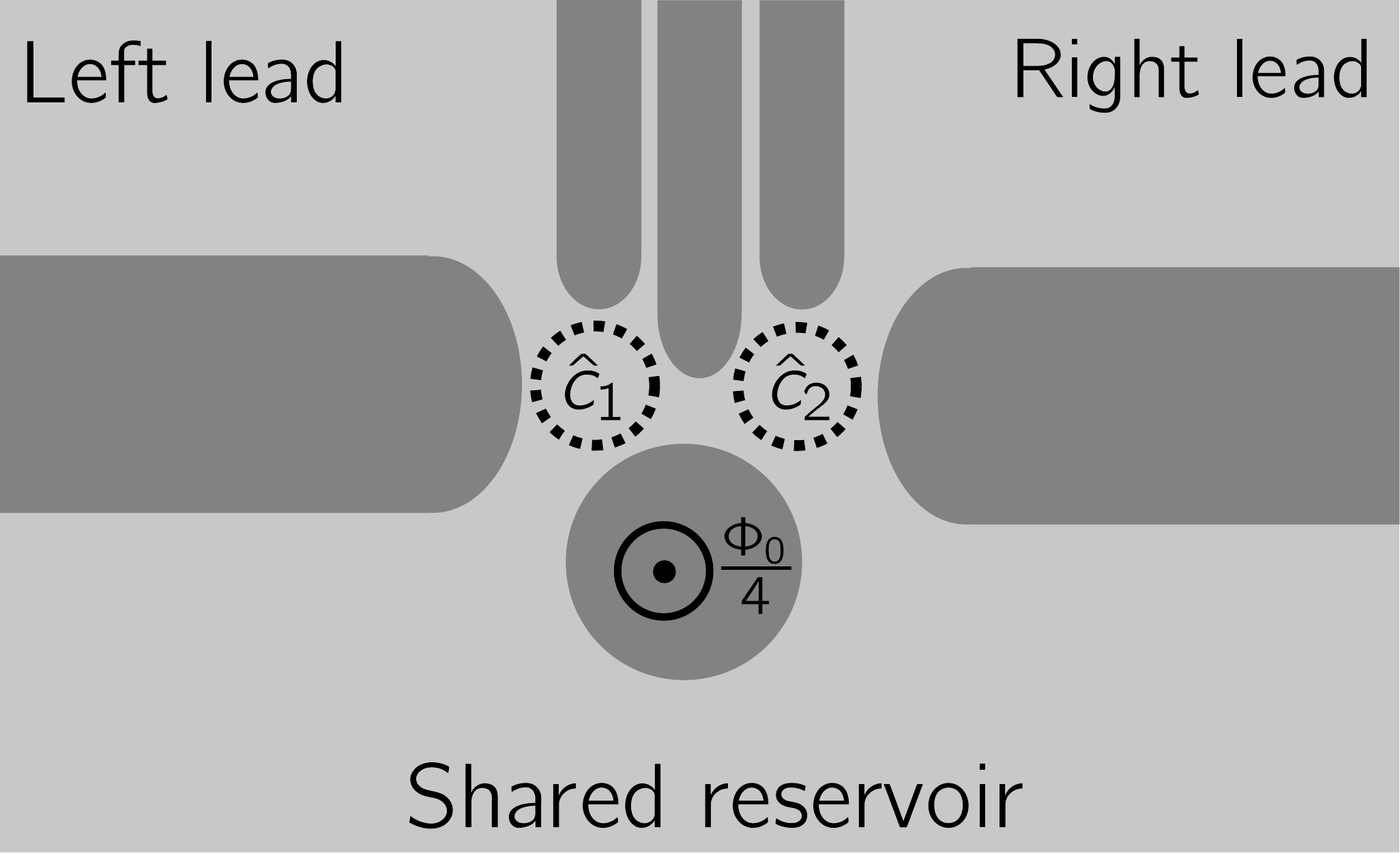}
	\caption{A sketch showing a potential experimental implementation with a gated GaAs/AlGaAs heterostructure.
		Gates that expel the 2D electron gas are drawn in dark gray.
  	}
	\label{fig:experiment}
\end{figure}

Without the shared reservoir, \cref{eq:Hamiltonian} is the standard Hamiltonian for a serial DQD~\cite{Stafford1996,Wunsch2005,Tu2008,Schaller2014}. In contrast to previous work, we include a third, shared reservoir, which can be realized experimentally by tunnel-coupling both sites to a wire or a 2D electron gas parallel to the structure. We propose a specific experiment below (\cref{fig:experiment}).

Let us first explore the mechanism for directionality within the quantum master equation (QME).
It is derived assuming the system is weakly coupled to its reservoirs and the Born-Markov approximation is valid~\cite{Breuer2002}.
The QME takes the Lindblad form (derivation in \cref{app:QME_derivation})
\begin{equation}
	\begin{aligned}
		\dot\dens_S=-i[\tilde H_{\text{sys}},\dens_S]+\sum_j[\gamma_{j}^-\D(\hat c_j)+\gamma_{j}^+\D(\hat c_j\dagg)]\dens_S\\
		+[\gamma_B^+\D(\hat c_1\dagg+\hat c_2\dagg)+\gamma_B^-\D(\hat c_1+\hat c_2)]\dens_S,
	\end{aligned}
	\label{eq:full_QME}
\end{equation}
with 
\begin{subequations}
	\begin{align}
		\tilde H_{\text{sys}}&=\tilde\eps(\hat n_1+\hat n_2)+\tfrac{\tilde\delta}{2}(\hat n_1-\hat n_2)+(\tilde\lambda \hat c_1\dagg \hat c_2+\text{H.c.}),\\
		\gamma_\alpha^+&=\Gamma_\alpha f(\eps-\mu_\alpha),\qquad
		\gamma_\alpha^- =\Gamma_\alpha [1-f(\eps-\mu_\alpha)]\label{eq:gamma},
	\end{align}
\end{subequations}
where the tilde denotes that the parameters have been renormalized by the self-energy due to the reservoirs, and $\tilde\eps\equiv(\tilde \eps_1+\tilde \eps_2)/2$, $\tilde\delta\equiv\tilde\eps_1-\tilde\eps_2$.
In the remainder of this article we will drop the tilde again.
The index $\alpha$ runs over $(1,2,B)$. The dissipation rates depend on the reservoir density of states at energy $\eps$ and the coupling amplitudes, which has been combined into the overall rate $\Gamma_\alpha$, as detailed in \cref{app:QME_derivation}.
$f(\eps)=\{\exp[ \eps/(k_BT)]+1\}^{-1}$ is the Fermi-Dirac distribution.
We assume all reservoirs to be at the same temperature, but allow the chemical potential to vary between the reservoirs, as they will be set by the applied voltages.

There is extensive literature about whether the QME should be derived with respect to local degrees of freedom or with respect to global energy eigenstates of the system \cite{Levy2014,Mitchison2017,Hofer2017a}.
In thermodynamic equilibrium, global dissipators tend to be more accurate, but in out-of-equilibrium situations, it has been shown that local dissipators model transport behavior more accurately~\cite{Hofer2017a}, which is why we have employed local dissipators here. In order to show that they do indeed capture the appropriate physics, we compare to the exact solution for reservoirs with infinite bandwidth below.

\begin{figure*}[t]
	\centering
	\includegraphics[width=\linewidth]{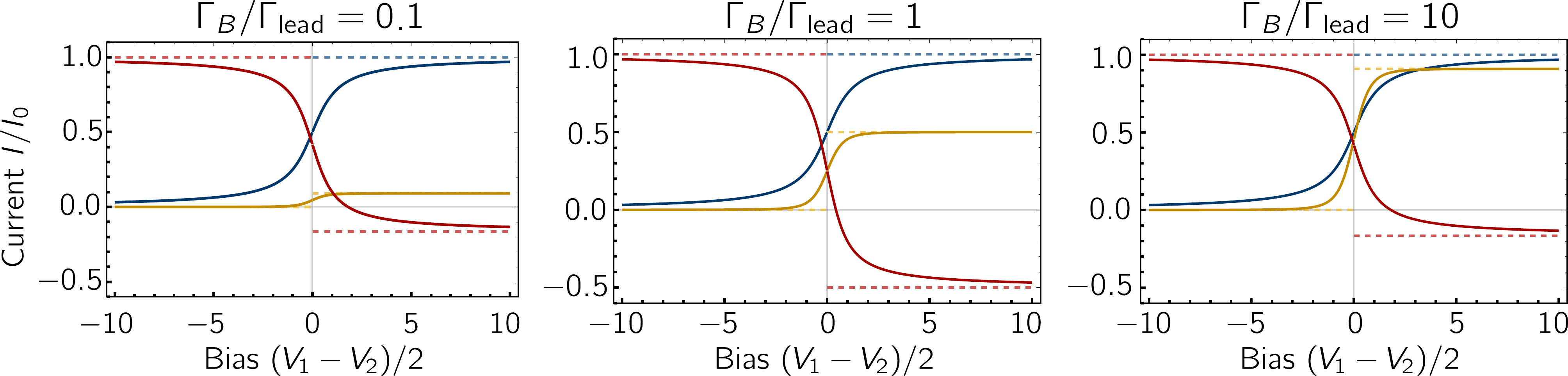}
	\caption{We plot the currents $\langle\hat I_1\rangle$ (blue), $\langle\hat I_{12}\rangle $ (yellow), $\langle\hat I_2\rangle $ (red) for strong coupling [\cref{eq:strong_I}, in solid, dark] and weak coupling (dotted, light), at zero temperature, as a function of the bias, where $V_1\equiv2(\mu_1-\eps)/(\Gamma_B+\Gamma_{\text{lead}})=-V_2$, for weak (left), intermediate (middle) and strong (right) inter-dot coupling relative to the coupling to the leads. The currents are plotted in units of $I_0=\Gamma_B\Gamma_{\text{lead}}/(\Gamma_B+\Gamma_{\text{lead}})$.
	In reverse bias, current from lead 2 flows into the shared reservoir, but current never flows into lead 1 and both $\langle\hat I_1\rangle $ and $\langle\hat I_{12}\rangle $ go to zero. 
	In forward (positive) bias, current flows from lead 1 to 2, but the current into lead 2 is at most half of the current leaving lead 1, which happens in the ``impedance-matched'' case where the inter-dot coupling rate $2|\lambda|=\Gamma_B$ equals the lead coupling rate $\Gamma_{\text{lead}}$. As the asymmetry in $\Gamma_B/\Gamma_{\text{lead}}$ grows, more electrons get directed into the shared reservoir [cf.~\cref{eq:strong_I}].
}
\label{fig:IV}
\end{figure*}

\section{Directionality}
Consider the equation of motion for the expectation value of the number of electrons on site 1,
$\hat n_1$, derived from \cref{eq:full_QME}
\begin{equation}
	\begin{aligned}
		\frac{d}{dt}\langle\hat n_1\rangle=&-(\Gamma_1+\Gamma_B)\langle\hat n_1\rangle
		-i\langle\lambda \hat c_1\dagg \hat c_2-\lambda^*\hat c_2\dagg \hat c_1\rangle\\
		&-\frac{\Gamma_B}{2}\langle\hat c_1\dagg \hat c_2+\hat c_2\dagg \hat c_1\rangle+(\gamma_B^++\gamma_1^+).
		\label{eq:n1_eom}
	\end{aligned}
\end{equation}
The terms in this equation describe (in order): loss of electrons into two reservoirs, \emph{coherent} tunneling of electrons between the two sites, \emph{dissipative} coupling arising from the non-local dissipator, and a constant rate of fermions added from the reservoirs.
The term $-i\langle\lambda \hat c_1\dagg \hat c_2-\lambda^*\hat c_2\dagg \hat c_1\rangle $ is the current between the two sites. It is canceled by the succeeding term in \cref{eq:n1_eom} if
\begin{equation}
	\lambda=i\Gamma_B/2
	\label{eq:directionality_condition}
\end{equation}
which causes destructive interference between the coherent and the dissipative process~\cite{Metelmann2015}.
This choice for $\lambda$, which we adopt for the rest of the article, makes $\langle\hat n_1\rangle $ independent of site 2, which is the essence of isolation.
Crucially, the same is not true for site 2, as we have
\begin{equation}
	\begin{aligned}
		\frac{d}{dt}\langle\hat n_2\rangle&=-(\Gamma_2+\Gamma_B)\langle\hat n_1\rangle
		+i\langle\lambda \hat c_1\dagg \hat c_2-\lambda^*\hat c_2\dagg \hat c_1\rangle\\
		&-\frac{\Gamma_B}{2}\langle\hat c_1\dagg \hat c_2+\text{H.c.}\rangle+(\gamma_B^++\gamma_2^+),
	\end{aligned}
\end{equation}
such that for our choice [\cref{eq:directionality_condition}] the current from site 1 to site 2 is \emph{enhanced}.
Mathematically, this happens because the phase in the coherent interaction is conjugated ($\lambda^*\hat c_2\dagg \hat c_1$) when exchanging 1 and 2, whereas the dissipator [$\D(\hat c_1+\hat c_2)$] is symmetric. 

While the QME enables a simple analysis, we gain confidence in our result by deriving the exact solution directly from the equations of motion, which is also valid for strong coupling.
Using the Laplace transform $\tilde c(z)\equiv\int_0^\infty\exp(-zt)c(t)dt$ allows us to write the equations of motion as algebraic ones
\begin{equation}
	\begin{aligned}
		&\mat{z+i\eps_1+i\Sigma_1(z) &
		i\lambda+\sum_k\frac{G_{k,1}G_{k,2}}{z+i\eps_{k,B}}\\
		i\lambda^*+\sum_k\frac{G_{k,2}G_{k,1}}{z+i\eps_{k,B}}&
		z+i\eps_2+i\Sigma_2(z)}
		\mat{\tilde c_1(z)\\\tilde c_2(z)}\\
		&=\mat{\hat c_1(0)+\sum_k\frac{iG_{k,1}}{z+i\eps_{k,B}}\hat b_{k,B}(0)+\sum_k\frac{iJ_{k,1}}{z+i\eps_{k,1}}\hat b_{k,1}(0)\\
		\hat c_2(0)+\sum_k\frac{iG_{k,2}}{z+i\eps_{k,B}}\hat b_{k,B}(0)+\sum_k\frac{iJ_{k,2}}{z+i\eps_{k,2}}\hat b_{k,2}(0)}
	\end{aligned}
	\label{eq:laplace_eom}
\end{equation}
with
\begin{equation}
	\begin{aligned}
		i\Sigma_j(z)&=\sum_k\left(\frac{G_{k,j}^2}{z+i\eps_{k,B}}+\frac{J_{k,j}^2}{z+i\eps_{k,j}}\right)\\
		&\to \int\frac{d\omega}{2\pi}\left( \frac{\Gamma_{j,B}(\omega)}{z+i\omega}+\frac{\Gamma_j(\omega)}{z+i\omega} \right).
	\end{aligned}
	\label{eq:self-energy}
\end{equation}
The matrix on the left-hand side of \cref{eq:laplace_eom} describes similar physical effects as the QME.
$\Sigma_i(z)$ is a complex self-energy induced by the coupling to the two reservoirs, which describes loss (imaginary part) and renormalization of the energy (real part). 
The inter-dot coupling $\lambda$ is also modified by an equivalent term, which captures the interference of coherent and dissipative coupling. Finally, the right-hand side of \cref{eq:laplace_eom} contains the initial state of the system. The correlators between the reservoir modes contain information about chemical potential and temperature of the reservoir.

For a dense set of reservoir modes, we can replace the sums over energy eigenstates (denoted symbolically by $\sum_k$), as done in \cref{eq:self-energy}.
In order to match the exact solution to the QME, we choose the reservoir spectral density to be flat, i.e., $\Gamma_\alpha(\omega)=\Gamma_\alpha$.
Assuming for simplicity that $\Gamma_{1,B}=\Gamma_{2,B}\equiv\Gamma_B$ (full solution in \cref{app:laplace}), directionality is attained again for $\lambda=i\Gamma_B/2$, in agreement with \cref{eq:directionality_condition}.
Furthermore, the fact that this effect occurs in the equations of motion for the operators $c_1, c_2$ [\cref{eq:laplace_eom}] is clear evidence that directionality arises due to interference.

\section{Currents}
Ultimately, the relevant quantities in experiment are the currents between the sites and through the leads. We derive them below for both the QME and the exact solution.

Together with the equation of motion for the expectation value of the inter-dot current operator $\hat I_{12}=-\Gamma_B(\hat c_1\dagg \hat c_2+\hat c_2\dagg \hat c_1)/2$
the QME yields a closed system of equations which is solved to obtain the steady-state expectation value (cf.~\cref{app:QME_solution})
\begin{equation}
	\langle\hat I_{12}\rangle=
	\frac{\Gamma_B^2\Gamma_{\text{lead}}[f(\eps-\mu_1)-f(\eps-\mu_B)]}{(\Gamma_B+\Gamma_{\text{lead}})^2+\delta^2},
  	\label{eq:I12QME}
\end{equation}
where we have set $\Gamma_i\equiv\Gamma_{\text{lead}}$ for simplicity.
\Cref{eq:I12QME} is a key result of our analysis.
In order to obtain fully directional transport we need $\gamma_B^+=0$, attained for $\eps-\mu_B\gg k_BT$, such that electrons from the shared reservoir do not enter the system.
In this case, the current is always non-negative, the hallmark of directional transport.
This is the regime we consider in the rest of the paper.

In \cref{eq:I12QME}, $\delta$ is the energy difference between the two sites.
If it is large compared to the dissipation rates, the two fermionic modes do not overlap, and current is suppressed.
If $\delta$ is negligible, and for strong bias ($\mu_1-\eps\gg k_BT$, such that $\gamma_1^+=\Gamma_{\text{lead}})$,
we have $\langle\hat I_{12}\rangle \approx \Gamma_B^2\Gamma_{\text{lead}}/(\Gamma_B+\Gamma_{\text{lead}})^2$, 
and we identify two limits.
If $\Gamma_B\gg\Gamma_{\text{lead}}$, inter-dot coupling is large compared to dot-lead coupling, 
and the current is dominated by the rate at which electrons are added: $\langle\hat I_{12}\rangle\approx\Gamma_{\text{lead}}$.
Conversely, if $\Gamma_B\ll\Gamma_{\text{lead}}$, the current is dominated by the rate at which electrons are shuttled from site 1 to 2:
$\langle\hat I_{12}\rangle\approx\Gamma_B^2/\Gamma_{\text{lead}}$.

\begin{figure}[t]
	\includegraphics[width=.8\linewidth]{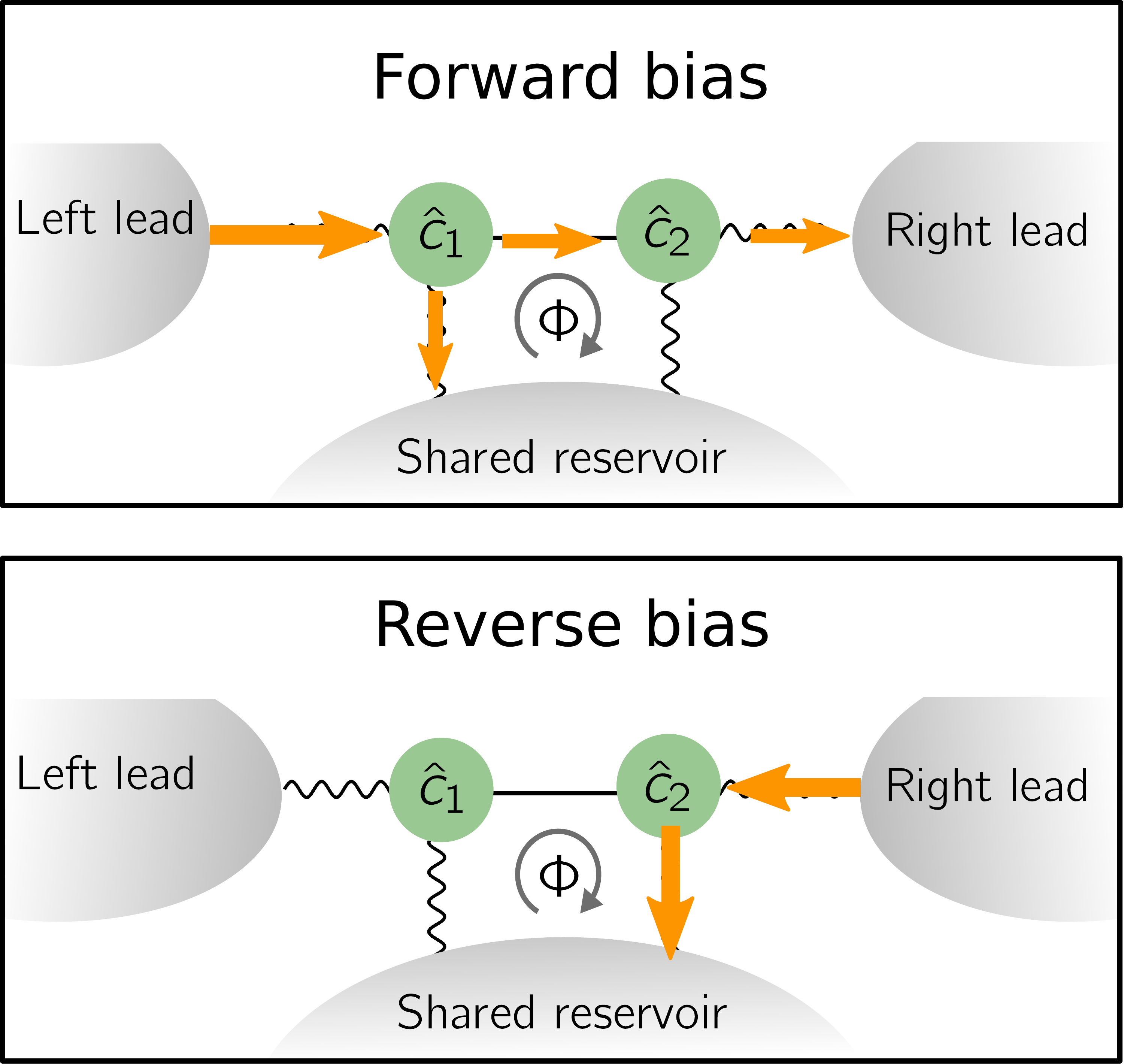}
	\caption{This plot displays schematically how currents flow in the case of forward and reverse bias,
		for a DQD with impedance-matched inter-dot and dot-lead coupling rates $\Gamma_B=\Gamma_{\text{lead}}$, and in the directional regime $\lambda=i\Gamma_B/2$.
		In reverse bias, the whole current from the lead is absorbed in the shared reservoir, and no current arrives in the left lead. 
		On the other hand, in forward bias, half of the current is absorbed by the shared reservoir, and the other half is transmitted, which can be seen in \cref{eq:strong_I}.
		}
	\label{fig:dqd_current}
\end{figure}

Intriguingly, current from the shared reservoir reduces $\langle\hat I_{12}\rangle$.
While it could seem surprising or worrying that electrons seemingly flow against directionality,
it is a natural consequence of the fact that the directionality originates from interference.
Electrons on site 2 have zero amplitude of traveling to site 1, but this is not true for electrons from the shared reservoir, which are added in a superposition on sites 1 and 2.
Despite this, our system is not a circulator, as can be seen from the asymmetry between the currents from the three terminals (cf.~\cref{app:QME_solution}).

\begin{figure}[t]
	\centering
	\includegraphics[width=.75\linewidth]{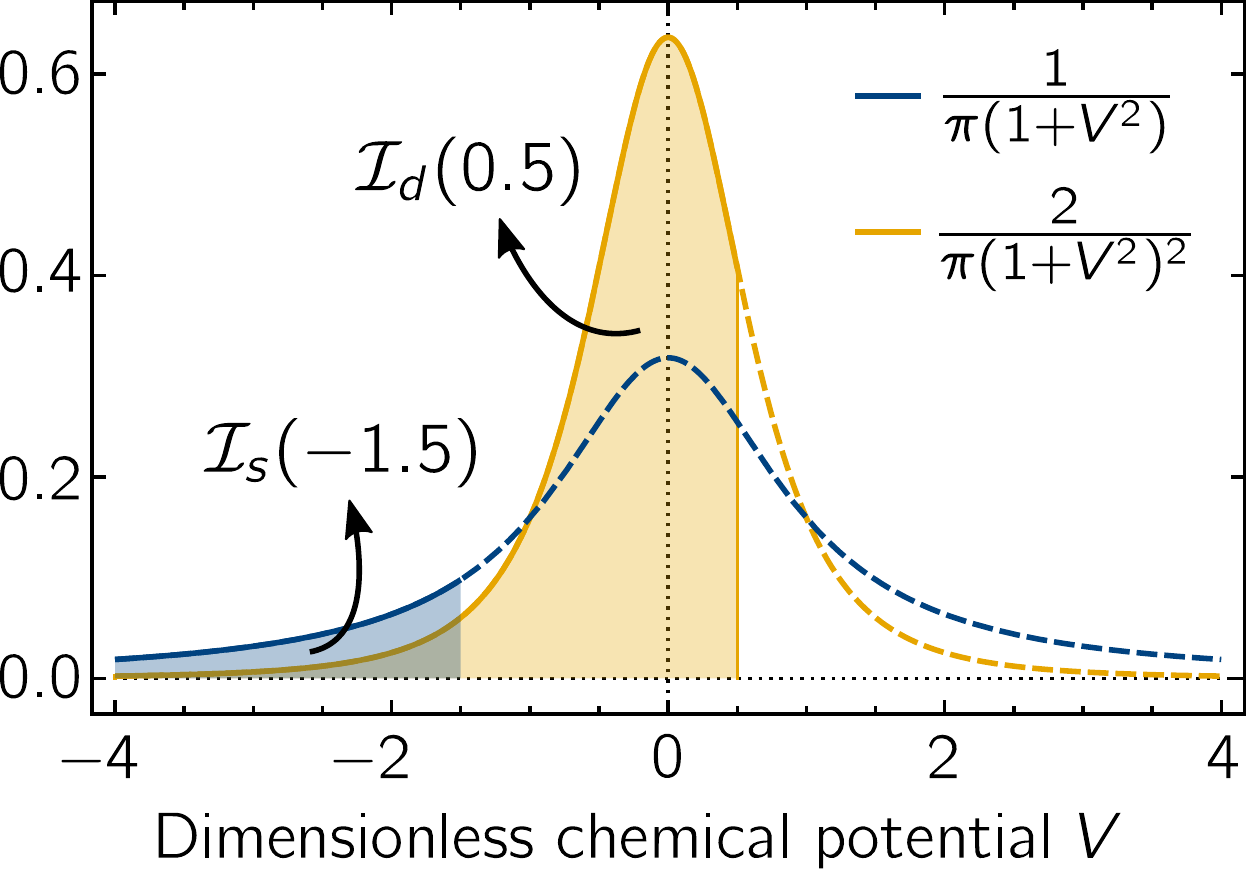}
	\caption{$\I_s$ ($\I_d$) is the integral over a normalized Lorentzian (squared) from $-\infty$ to the normalized chemical potential $V$.
	}
	\label{fig:lorentzian}
\end{figure}

To verify \cref{eq:I12QME}, we present the exact solution obtained from \cref{eq:laplace_eom}, and compare it to the QME in \cref{fig:IV}.
Inverting the Laplace transform yields the real-time solution for all operators, whose correlators converge to stationary values in the long-time limit, which are generically expressed as integrals over all energies.
At zero temperature, the inter-dot current $\langle\hat\I_{12}\rangle$ and
the current leaving lead $i=1,2$, $\langle\hat\I_i\rangle$, become
\begin{subequations}
	\begin{align}
  		\langle\hat{\I}_{1}\rangle&=I_0\I_s(V_1),\quad \langle\hat{\I}_{12}\rangle=
		\frac{I_0\Gamma_{\text{lead}}}{\Gamma_B+\Gamma_{\text{lead}}}\I_d(V_1),
		\label{eq:I12}\\
  		\langle\hat{\I}_{2}\rangle&=I_0\I_s(V_2)
  		-\frac{2\Gamma_B}{\Gamma_B+\Gamma_{\text{lead}}}\langle\hat\I_{12}\rangle.
	\end{align}
  	\label{eq:strong_I}
\end{subequations}
where the scaled chemical potential $V_\alpha\equiv2(\mu_\alpha-\eps)/(\Gamma_B+\Gamma_{\text{lead}})$, 
and we have defined $I_0\equiv(\Gamma_B\Gamma_{\text{lead}})/(\Gamma_B+\Gamma_{\text{lead}})$ and the currents through a single ($s$) and double ($d$) dot
\begin{equation}
  	\I_s(V)=\frac{1}{2}+\frac{\tan^{-1}(V)}{\pi},\,
  	\I_d(V)=\I_s(V)+\frac{V}{\pi(1+V^2)},
	\label{eq:I_defn}
\end{equation}
which are the integral over a Lorentzian and the square of a Lorentzian, respectively (illustrated in \cref{fig:lorentzian}).
Alternatively, $V$ can be considered a scaled voltage with respect to the energy of the site $\eps$.

It is known that $I_0$ is the maximum current through a mode (per spin)~\cite{Datta1997} and that current through a mode is proportional to the area under the lineshape up to the chemical potential~\cite{Datta1997}.
At finite temperature, $\I_{s,d}$ are modified, but \cref{eq:strong_I} remains unchanged.
We distinguish expectation values in the exact solution by using a calligraphic $\I$, even though the current operator is the same in both cases.
The QME result (at $\delta=0$) can be obtained from \cref{eq:strong_I} by replacing
\begin{equation}
	\I_{s,d}(V)\quad\rightarrow\quad f(\mu-\eps).
	\label{eq:replacement}
\end{equation}
Essentially, the weak-coupling QME neglects the finite width of the modes.

We plot the current-voltage characteristics for symmetric bias, $V_2=-V_1$, zero temperature, and $V_B\to-\infty$ for both solutions in \cref{fig:IV}. The current leaving the first lead coincides with current through a single dot~\cite{Schaller2014}, reflected in $\I_s$.
The second lead additionally receives current from the first lead, which passes through both dots and hence has a characteristic given by $\I_d$.

The current is clearly directional, in the sense that current never enters the first lead, even in reverse bias. However, some current is directed into the shared reservoir. In the ideal case, where inter-dot coupling and dot-lead coupling rates are matched, $\Gamma_B=\Gamma_{\text{lead}}$, and for $V_1\gg1$, $\langle\hat{\I}_1\rangle\to I_0$, whereas $\langle\hat{\I}_{2}\rangle\to-I_0/2$ and half of the current flows into the shared reservoir, as shown in \cref{fig:dqd_current}.
Away from that point the amount of current lost increases steadily [cf.~\cref{eq:strong_I,fig:IV}].

\section{Experimental implementation}
Our proposal can be realized in gated GaAs/AlGaAs heterostructures, a well-established platform for QDs~\cite{Wiel2002,Hanson2007}, where related systems are a reality~\cite{Schuster1997,Holleitner2001,Avinun-Kalish2005}.
Directionality requires finely tuned coupling rates, which are achievable in current experiments~\cite{Elzerman2003,Koppens2006,Baart2016}. Island gates with magnetic flux have been implemented before~\cite{Schuster1997,Avinun-Kalish2005}.
A magnetic flux of $\Phi_0/4$ threading an area of $0.01\,\mu$m$^2$---a typical scale for experiments~\cite{Schuster1997,Holleitner2001,Avinun-Kalish2005,Hanson2007,Braakman2013}---requires a magnetic field of approximately $50\,$mT, which is routinely achieved.
If not confined to the island, this magnetic field simultaneously serves to spin-polarize the dots.

\section{Discussion}
One important open question concerns the effects of structure in the various reservoirs on the directionality properties. In \cref{eq:laplace_eom} we see that isolation occurs when $i\lambda=-\int(d\omega/2\pi)\Gamma_B(\omega)/(z+i\omega)$, independent of the leads, such that we can confidently conclude that structure in the leads does not impact directionality---though clearly a finite bandwidth of the shared reservoir does.
We expect isolation to work well when the characteristic frequency range $\Delta\Gamma$ over which the reservoir density of states changes is large compared to the width of the system modes, $\Delta\Gamma\gg\Gamma_B,\Gamma_{\text{lead}}$.
Several numerical approaches have been developed to tackle non-Markovian reservoirs~\cite{DeVega2017}.
Approaches that extend the mode space of the quantum system~\cite{Strasberg2017} might be particularly suitable.

Decoherence processes that couple to the number operator, such as the phonon reservoir or the Coulomb interaction, do not affect the mechanism for directionality,
since the equation of motion for $\hat n_i$ does not change when dissipators such as $\D(\hat n_1)$ or $\D(\hat n_1\pm \hat n_2)$ are added.
The current between the dots, however, is reduced as the coherence between the sites is lost, akin to a quantum Zeno effect. 

A more realistic double quantum dot model might include a non-linear Coulomb-repulsion term $\xi \hat c_1\dagg \hat c_1\hat c_2\dagg \hat c_2$ in ~\cref{eq:Hamiltonian}.
It is not immediately clear how such a term modifies directionality. While it precludes the straightforward solution via equations of motion, it commutes with $\hat n_1,\hat n_2,\hat I_{12}$, and thus does not alter the QME result, but the QME derived here might cease to be applicable.

Finally, since the equations are linear, the QME result can easily be generalized to $\lambda\neq i\Gamma_B/2$, which could become relevant for experiment.

\section{Conclusion}
We have introduced fermionic reservoir engineering in DQDs and shown that a third reservoir shared between both sites of a serial DQD leads to current rectification.
The effect is robust to various sources of decoherence and is observable with current quantum dot technology.

\begin{acknowledgments}
	We are grateful to Mark Buitelaar, Nigel Cooper, Fernando Gonzales-Zalba, James Haigh, and Alessandro Rossi for insightful discussions and helpful comments.
	DM acknowledges support by the UK Engineering and Physical Sciences Research Council (EPSRC) under Grant No.\ EP/M506485/1.
	AN holds a University Research Fellowship from the Royal Society and acknowledges support from the Winton Programme for the Physics of Sustainability
	and the European Union's Horizon 2020 research and innovation programme under grant agreement No 732894 (FET Proactive HOT).
\end{acknowledgments}

\appendix
\section{Derivation of dissipators in the quantum master equation}\label{app:QME_derivation}
Assuming the Born-Markov approximation, the equation of motion for the density matrix is given by the QME~\cite{Breuer2002}
\begin{equation}
	\frac{d}{dt}\dens_S(t)=-\int_0^\infty ds\,\tr_B\left[ \hat H_I(t),\left[ \hat H_I(t-s),\dens_S(t)\otimes\dens_B \right] \right],
\end{equation}
where $\hat H_I(t)$ is the interaction-picture Hamiltonian for the interaction with the reservoirs.
Here we take the bare Hamiltonian $\hat H_0=\eps(\hat n_1+\hat n_2)+\hat H_{\text{res}}$ and leave out the energy splitting $\hat H_\delta=(\delta/2)(\hat n_1-\hat n_2)$ as well as inter-dot coupling $\hat H_\lambda=\lambda\hat c_1\dagg\hat c_2+\lambda^*\hat c_2\dagg\hat c_1$,
such that the interaction-picture Hamiltonian for the system-reservoir coupling becomes
\begin{equation}
	\begin{aligned}
		\hat H_I(t)&=-\sum_k\sum_{j=1}^2\hat c_j\dagg e^{i\eps t} \left(G_{k,j}\hat b_{k,B}e^{ikx_i-i\eps_{k,B}t}\right.\\
		&\qquad+\left.J_{k,j}\hat b_{k,j}e^{-i\eps_{k,j}t}\right)+\text{H.c.}
	\end{aligned}
	\label{eq:single site interaction picture Hamiltonian}
\end{equation}
$\dens_B$ is the reservoir density matrix, which remains unchanged over time (Born approximation).
Here, we will assume it to be thermal, with a given chemical potential, such that the occupation of each mode is governed by the Fermi-Dirac distribution.
If there are no correlations between the reservoirs, we can treat them separately.

The part that couples site $j$ to lead $j$
\begin{equation}
	\hat H_{I,j}(t)=-\sum_k\hat c_j\dagg e^{i\eps t}J_{k,j}\hat b_{k,j}e^{-i\eps_{k}t}+\text{H.c.}
\end{equation}
leads to a contribution to the QME
\begin{equation}
	\begin{aligned}
		\dot\dens_S&=|J_{k_0,j}|^22\pi\nu_j(\eps)\left[ (1-f_j(\eps))\D(\hat c_j)+f_j(\eps)\D(\hat c_j\dagg) \right]\dens_S\\
		&\qquad-i \Re[\Sigma_j][\hat c_j\dagg \hat c_j,\dens_S]+\cdots,
	\end{aligned}
	\label{eq:lead_QME}
\end{equation}
where $k_0$ is the wavevector at which $\eps_{k_0}=\eps$, and $f_j(\eps)=\{1+\exp[(\eps-\mu_j)/(k_BT_j)]\}^{-1}$.
The first term corresponds to incoherent particle loss or gain, depending on temperature, chemical potential, and the energy of the site.
The second term renormalizes the energy of the site, given by the self-energy $\Re[\Sigma_j]\equiv \sum_k|J_{k,j}|^2\P(1/(\eps_k-\eps))$, where $\P$ denotes the principal part and the dots denote that this is only part of the equation of motion for $\dens_S$.

We repeat the analysis for the shared reservoir
\begin{equation}
	\hat H_{I,B}(t)=-\sum_j\hat c_j\dagg G_{k,j} \hat b_{k,B}e^{i\eps t+ikx_j-i\eps_{k,B}t}+\text{H.c.}
\end{equation}
Going through the same procedure as before, we arrive at
\begin{equation}
	\begin{aligned}
		\dot\dens_S&=-\sum_{k,i,j}\left\{ (1-f_B(\eps_k))\left[ G_{ij}(k,t)(\hat c_i\dagg \hat c_j\dens-\hat c_j\dens \hat c_i\dagg)\right.\right.\\
		&+\left.G_{ij}^*(k,t)(-\hat c_i\dens \hat c_j\dagg+\dens \hat c_j\dagg \hat c_i)\right]\\
		&+f_B(\eps_k)\left[ G_{ij}(k,t)(-\hat c_i\dagg\dens \hat c_j+\dens \hat c_j\hat c_i\dagg)\right.\\
		&+\left.\left.G_{ij}^*(k,t)(\hat c_i\hat c_j\dagg\dens-\hat c_j\dagg\dens \hat c_i) \right] \right\}+\cdots,
	\end{aligned}
\end{equation}
with
\begin{equation}
	G_{ij}(k,t)=G_{k,i}G_{k,j}^*e^{ik(x_i-x_j)}\left(\P\frac{-1}{i(\eps_k-\eps)}+\pi\delta(\eps_k-\eps) \right).
\end{equation}

Rearranging yields
\begin{equation}
	\begin{aligned}
		\dot\dens_S&=\sum_k2\pi\delta(\eps_k-\eps)\Big[ (1-f_B(\eps_k))\D(\hat z_k)\dens_S\\
		&+\left.f_B(\eps_k)\D(\hat z_k\dagg)\dens_S \right]
		+\P\frac{1}{i(\eps_k-\eps)}[\hat z_k\dagg \hat z_k,\dens_S]+\cdots,
	\end{aligned}
	\label{eq:shared_QME}
\end{equation}
with 
\begin{equation}
	\hat z_k\equiv G_{k,1}^*e^{-ikx_1}\hat c_1+G_{k,2}^*e^{-ikx_2}\hat c_2.
	\label{eq:zk}
\end{equation}
The first term can be evaluated, due to the presence of the delta function, and the second can be written as an effective Hamiltonian.
In order to evaluate the delta function, we assume
\begin{itemize}
	\item that the reservoir dispersion relation is symmetric at the energy $\eps$, i.e., that $\eps_{-k_0}=\eps_{k_0}$.
		This can be tuned with a current through the reservoir, which is another way to obtain an overall complex phase, such that directionality may be obtained without a magnetic field.
		Note that the factor of 2 disappears because the density of states includes the states at positive and negative wavevector, which we have to write out explicitly.
	\item symmetric coupling $G_{k,i}=G_{k}$ and choose $G_{k_0}\in\R$. Any phase can be incorporated into inter-dot coupling $\lambda$.
\end{itemize}
Simplifying, and including the two individual leads, we arrive at the QME in Lindblad form
\begin{equation}
	\begin{aligned}
		\dot\dens_S&=-i[\tilde H_{\text{sys}},\dens_S]+\sum_j[\gamma_{j}^-\D(\hat c_j)+\gamma_j^+\D(\hat c_j\dagg)]\dens_S\\
		&+[\gamma_B^+\D(\hat c_1\dagg+\hat c_2\dagg)+\gamma_B^-\D(\hat c_1+\hat c_2)]\dens_S,
	\end{aligned}
	\label{eq:full_QME_app}
\end{equation}
with
\begin{subequations}
	\begin{align}
		\tilde H_{\text{sys}}&=\hat H_0+\hat H_\delta+\hat H_{\lambda}+\hat H_{\text{self-energies}},\\
		\gamma_B^+ &= 2\pi G_{k_0}^2\nu_B(\eps)f_B(\eps)\cos[k_0(x_1-x_2)],\label{eq:gamma_app}\\
		\gamma_B^- &= 2\pi G_{k_0}^2\nu_B(\eps)[1-f_B(\eps)]\cos[k_0(x_1-x_2)],\\
		\gamma_j^+ &= 2\pi J_{k_0,j}\nu_j(\eps)f_j(\eps)\\
		&+2\pi\nu_B(\eps)G_{k_0}^2f_B(\eps)\{1-\cos[k_0(x_1-x_2)]\},\nonumber\\
		\gamma_j^- &= 2\pi J_{k_0,j}\nu_j(\eps)[1-f_j(\eps)]\label{eq:gamma-_app}\\
		&+2\pi\nu_B(\eps)G_{k_0}^2[1-f_B(\eps)]\{1-\cos[k_0(x_1-x_2)]\}\nonumber
,
	\end{align}
\end{subequations}
and where $H_{\text{self-energies}}$ is the sum of the terms in \cref{eq:lead_QME,eq:shared_QME}.
In order to derive the exact form of the dissipation rates, we started by assuming the coupling rates to the shared reservoir take the form $G_{k}e^{i\eps_{k,i}x}$.
This specific form is unlikely to be present in a realistic system. However, the resulting $\cos(\phi)$ term can be used to parametrize the imbalance between the reservoir couplings, with $\phi$ varying from $-\pi/2$ to $\pi/2$.
This will modify the precise form of the rates, but not change the physics fundamentally.
These are subtleties that we do not wish to address in this paper, and hence we set $\cos[k_0(x_1-x_2)]=1$.

With this choice, we arrive at the expressions in the main text
\begin{subequations}
	\begin{align}
		\gamma_\alpha^+ &=\Gamma_\alpha f_\alpha(\eps),\qquad
		\gamma_\alpha^- =\Gamma_\alpha[1-f_\alpha(\eps)],
	\end{align}
\end{subequations}
where $\alpha\in\{1,2,B\}$, and $\Gamma_\alpha\equiv\gamma_\alpha^++\gamma_\alpha^-$.
Note that if the temperature is equal across all reservoirs, we can write $f_\alpha(\eps)=f(\eps-\mu_\alpha)$,
with $f(\eps)=\{1+\exp[\eps/(k_BT)]\}^{-1}$, as is done in the main text.

\section{Solution of equations of motion}\label{app:QME_solution}
From the master equation, we derive the following equations of motion
\begin{subequations}
	\begin{align}
		\frac{d}{dt}\langle \hat n_1\rangle&=-(\Gamma_1+\Gamma_B)\langle \hat n_1\rangle
		-i\langle\lambda \hat c_1\dagg \hat c_2-\lambda^*\hat c_2\dagg \hat c_1\rangle\nonumber\\
		&-\frac{\Gamma_B}{2}\langle \hat c_1\dagg \hat c_2+\text{H.c.}\rangle+(\gamma_B^++\gamma_1^+),\\
		\frac{d}{dt}\langle \hat n_2\rangle&=-(\Gamma_2+\Gamma_B)\langle \hat n_2\rangle
		+ i\langle\lambda \hat c_1\dagg \hat c_2-\lambda^*\hat c_2\dagg \hat c_1\rangle\nonumber\\
		&-\frac{\Gamma_B}{2}\langle \hat c_1\dagg \hat c_2+\text{H.c.}\rangle+(\gamma_B^++\gamma_2^+),\\
		\frac{d}{dt}\langle \hat c_1\dagg \hat c_2\rangle&= (i\delta-\Gamma_y)\langle \hat c_1\dagg \hat c_2\rangle
		+i\lambda^*\langle \hat n_2-\hat n_1\rangle\nonumber\\
		&-\frac{\Gamma_B}{2}\langle \hat n_1+\hat n_2\rangle+\gamma_B^+,
	\end{align}
	\label{eq:QME_eoms}
\end{subequations}
having defined $\Gamma_B\equiv\gamma_B^++\gamma^-$, $\Gamma_j=\gamma_j^++\gamma_j^-$,
and $\Gamma_y\equiv\Gamma_B+(\Gamma_1+\Gamma_2)/2$.
Setting $\lambda=i\Gamma_B/2$, we arrive at
\begin{subequations}
	\begin{align}
		\frac{d\langle \hat n_1\rangle }{dt}&=-(\Gamma_1+\Gamma_B)\langle \hat n_1\rangle +\gamma_1^++\gamma_B^+,\\
		\frac{d\langle \hat n_2\rangle }{dt}&=-(\Gamma_2+\Gamma_B)\langle \hat n_2\rangle +\gamma_2^++\gamma_B^+\nonumber\\
		&-\Gamma_B\langle \hat c_1\dagg \hat c_2+\hat c_2\dagg \hat c_1\rangle,\\
		\frac{d}{dt}\langle \hat c_1\dagg \hat c_2\rangle&=(i\delta-\Gamma_y)\langle \hat c_1\dagg \hat c_2\rangle-\Gamma_B\langle \hat n_1\rangle+\gamma_B^+ .
	\end{align}
	\label{eq:expectation_value_ODE}
\end{subequations}

The steady-state solution is obtained by setting Eqs.~\eqref{eq:expectation_value_ODE} to zero
\begin{subequations}
	\begin{align}
		\langle \hat n_1\rangle&=\frac{\gamma_1^++\gamma_B^+}{\Gamma_1+\Gamma_B},\\
		\langle \hat c_1\dagg \hat c_2\rangle&=\frac{\Gamma_1\gamma_B^+-\Gamma_B\gamma_1^+}{(\Gamma_1+\Gamma_B)(\Gamma_y-i\delta)},\\
		\langle \hat n_2\rangle&=\frac{\gamma_2^++\gamma_B^+}{\Gamma_2+\Gamma_B}+\frac{2\Gamma_y\Gamma_B(\Gamma_B\gamma_1^+-\Gamma_1\gamma_B^+)}{(\Gamma_1+\Gamma_B)(\Gamma_2+\Gamma_B)(\Gamma_y^2+\delta^2)}.
	\end{align}
\end{subequations}
In the limit considered in the main text, $\Gamma_i=\Gamma_{\text{lead}},\Gamma_{i,B}=\Gamma_B,\delta=0$, and zero temperature, these turn into
\begin{subequations}
	\begin{align}
		\langle \hat n_1\rangle&=\frac{\Gamma_{\text{lead}}\Theta(V_1)+\Gamma_B\Theta(V_B)}{\Gamma_{\text{lead}}+\Gamma_B},\\
		\langle \hat c_1\dagg \hat c_2\rangle&=\frac{\Gamma_{\text{lead}}\Gamma_B}{(\Gamma_{\text{lead}}+\Gamma_B)^2}\left[\Theta(V_B)-\Theta(V_1)\right],\\
		\langle \hat n_2\rangle&=\frac{\Gamma_{\text{lead}}\Theta(V_2)+\Gamma_B\Theta(V_B)}{\Gamma_{\text{lead}}+\Gamma_B}\nonumber\\
		&+\frac{2\Gamma_B^2\Gamma_{\text{lead}}}{(\Gamma_{\text{lead}}+\Gamma_B)^3}\left[\Theta(V_1)-\Theta(V_B)\right],
	\end{align}
	\label{eq:weak_coupling_occupations}
\end{subequations}
where $\Theta$ is the Heaviside step function.

\section{Lead currents and inter-dot current operator}\label{app:current_operator}
In order to find the current flowing from one site 1 to site 2, we consider the Heisenberg equation of motion for the number of particles at site 1 (in the absence of reservoirs)
\begin{equation}
	\dot{\hat n}_1=i[\hat H_S,\hat n_2]=-i(\lambda \hat c_1\dagg \hat c_2-\lambda^*\hat c_2\dagg \hat c_1)=\frac{\Gamma_B}{2}(\hat c_1\dagg \hat c_2+\hat c_2\dagg \hat c_1).
\end{equation}
We can interpret the RHS as the current from site 2 to 1 or as minus the current from site 1 to 2.
Its expectation value in the steady-state of the full model is
\begin{equation} 
	\begin{aligned}
		\langle \hat I_{12}\rangle&=-\frac{\Gamma_B}{2}\langle \hat c_1\dagg \hat c_2+\hat c_2\dagg \hat c_1\rangle
		=\frac{\Gamma_B\Gamma_y}{\Gamma_y^2+\delta^2}\frac{\Gamma_B\gamma_1^+-\Gamma_1\gamma_B^+}{\Gamma_{1}+\Gamma_B}\\
		&\xrightarrow{ \Gamma_i=\Gamma_{\text{lead}}}
		\Gamma_B^2\frac{\Gamma_B\gamma_1^+-\Gamma_{\text{lead}}\gamma_B^+}{(\Gamma_{\text{lead}}+\Gamma_B)^2+\delta^2}.
	\end{aligned}
	\label{eq:full_QME_current}
\end{equation}

The currents between the sites and the reservoirs have to be found in a slightly roundabout way. 
Considering again the equations of motion for the number of particles on site 1, we can write it as
\begin{equation}
	\frac{d}{dt}\langle \hat n_1\rangle =\gamma_1^+(1-\langle \hat n_1\rangle )-\gamma_1^-\langle \hat n_1\rangle -\Gamma_B\langle \hat n_1\rangle +\gamma_B^+.
\end{equation}
This form makes it clear that the current from the left lead to the first site is given by 
\begin{equation}
	\langle \hat I_{1}\rangle=\gamma_1^+(1-\langle \hat n_1\rangle )-\gamma_1^-\langle \hat n_1\rangle .
\end{equation}
Analogously we can find the current from the right lead onto site 2, $\langle \hat I_{2}\rangle$. Plugging in the solution above,
\begin{subequations}
	\begin{align}
		\langle \hat I_1\rangle &=\frac{\Gamma_B\gamma_1^+-\Gamma_1\gamma_B^+}{\Gamma_1+\Gamma_B},\\
		\langle \hat I_2\rangle &=\frac{\Gamma_B\gamma_2^+-\Gamma_2\gamma_B^+}{\Gamma_2+\Gamma_B}
		-\frac{2\Gamma_y\Gamma_2\Gamma_B(\Gamma_B\gamma_1^+-\Gamma_1\gamma_B^+)}{(\Gamma_1+\Gamma_B)(\Gamma_2+\Gamma_B)(\Gamma_y^2+\delta^2)}.
	\end{align}
\end{subequations}

\section{Exact solution through Laplace transform of equations of motion}\label{app:laplace}
We derive the following equations of motion from the Hamiltonian in the main text
\begin{subequations}
	\begin{align}
		\dot {\hat c}_1&=-i\eps_1\hat c_1-i\lambda   \hat c_2+i\sum_k\left( G_{k,1}\hat b_{k,B}+J_{k,1}\hat b_{k,1} \right),\\
		\dot {\hat c}_2&=-i\eps_2\hat c_2-i\lambda^* \hat c_1+i\sum_k\left( G_{k,2}\hat b_{k,B}+J_{k,2}\hat b_{k,2} \right),\\
		\dot {\hat b}_{k,B}&=-i\eps_{k,B}\hat b_{k,B}+iG_{k,1}\hat c_1+iG_{k,2}\hat c_2,\\
		\dot {\hat b}_{k,i}&=-i\eps_{k,i}\hat b_{k,i}+iJ_{k,i}\hat c_i.
	\end{align}
\end{subequations}
Through a Laplace transform $\tilde c_1(z)=\int_0^\infty dt\,\exp(-zt)\hat c_1(t)$, these equations can be turned into algebraic ones. Eliminating the reservoir modes
\begin{subequations}
	\begin{align}
		\tilde b_{k,B}(z)&=\frac{1}{z+i\eps_{k,B}}\left( \hat b_{k,B}(0)+iG_{k,1}\tilde c_1(z)+iG_{k,2}\tilde c_2(z) \right),\\
		\tilde b_{k,i}(z)&=\frac{1}{z+i\eps_{k,i}}\left( \hat b_{k,i}(0)+iJ_{k,i}\tilde c_i(z) \right),
	\end{align}
	\label{eq:reservoir_modes_laplace}
\end{subequations}
we arrive at
\begin{equation}
	\begin{aligned}
		&\mat{z+i\tilde\eps_1&
		i\lambda+\sum_k\frac{G_{k,1}G_{k,2}}{z+i\eps_{k,B}}\\
		i\lambda^*+\sum_k\frac{G_{k,2}G_{k,1}}{z+i\eps_{k,B}}&
		z+i\tilde\eps_2}
		\mat{\tilde c_1(z)\\\tilde c_2(z)}\\
		&=\mat{\hat c_1(0)+\sum_k\frac{iG_{k,1}}{z+i\eps_{k,B}}\hat b_{k,B}(0)+\sum_k\frac{iJ_{k,1}}{z+i\eps_{k,1}}\hat b_{k,1}(0)\\
		\hat c_2(0)+\sum_k\frac{iG_{k,2}}{z+i\eps_{k,B}}\hat b_{k,B}(0)+\sum_k\frac{iJ_{k,2}}{z+i\eps_{k,2}}\hat b_{k,2}(0)}\\
		&\qquad\equiv \mat{\tilde c_{1,\text{in}}(z)\\\tilde c_{2,\text{in}}(z)},
	\end{aligned}
	\label{eq:laplace_space}
\end{equation}
where the energy of the modes has been modified
\begin{equation}
	\tilde\eps_i\equiv\eps_i-i\sum_k\frac{G_{k,i}^2}{z+i\eps_{k,B}}-i\sum_k\frac{J_{k,i}^2}{z+i\eps_{k,i}}.
	\label{eq:modified_mode_energy}
\end{equation}
In order to make progress, we will have to make assumptions about the spectrum of reservoir modes. 
Here, we assume them to be dense (such that we have proper dissipation) and write
\begin{equation}
	\sum_{k}\frac{G_{k,i}^2}{z+i\eps_{k,B}}=\int \frac{d\omega}{2\pi}\frac{\Gamma_{i,B}(\omega)}{z+i\omega},
	\label{eq:dense_reservoir_spectrum}
\end{equation}
and 
\begin{equation}
	\sum_{k}\frac{J_{k,i}^2}{z+i\eps_{k,i}}=\int \frac{d\omega}{2\pi}\frac{\Gamma_{i}(\omega)}{z+i\omega}.
\end{equation}
We will further assume the tunneling rates to be Lorentzians $\Gamma(\omega)=\Gamma\delta^2/(\omega^2+\delta^2)$, and let the bandwidth $\delta\to\infty$. Non-Markovian effects can be included by keeping $\delta$ finite.
Together, these choices simplify \cref{eq:laplace_space} to
\begin{equation}
	\begin{aligned}
		&\mat{z+i\tilde\eps_1 & i\lambda+\frac{\sqrt{\Gamma_{1,B}\Gamma_{2,B}}}{2}\\
		i\lambda^*+\frac{\sqrt{\Gamma_{1,B}\Gamma_{2,B}}}{2} & z+i\tilde\eps_2}
		\mat{\tilde c_1(z)\\\tilde c_2(z)}\\
		&=\mat{\tilde c_{1,\text{in}}(z)\\ \tilde c_{2,\text{in}}(z)},
	\end{aligned}
	\label{eq:easy_laplace_space}
\end{equation}
where now 
\begin{equation}
	\tilde\eps_i=\eps_i-i\frac{\Gamma_1+\Gamma_{1,B}}{2}
	\label{eq:epstilde}
\end{equation}
We see that $2\lambda=i\sqrt{\Gamma_{1,B}\Gamma_{2,B}}$ leads to directional interaction (and that that the direction is flipped for the opposite phase). This choice makes the problem easier to solve as well. 
Here, isolation can be perfect due to the infinite bandwidth reservoirs. In a realistic setting, the bandwidth of the reservoir will limit the bandwidth of isolation.

We can express $\tilde c_i(z)$ in terms of the input operators by inverting the matrix
\begin{equation}
	\mat{\tilde c_1(z)\\\tilde c_2(z)}=\mat{\left(z+i\tilde\eps_1\right)^{-1} &0\\
	\frac{-\sqrt{\Gamma_{1,B}\Gamma_{2,B}}}{(z+i\tilde\eps_1)(z+i\tilde\eps_2)}
	& \left( z+i\tilde\eps_2 \right)^{-1}}
	\mat{\tilde c_{1,\text{in}}(z)\\ \tilde c_{2,\text{in}}(z)}.
\label{eq:laplace_solution}
\end{equation}

Due to the wide-band limit and directionality, the inverse Laplace transform can be found easily
\begin{widetext}
\begin{subequations}
	\begin{align}
		\tilde\chi_i(z)\equiv(z+i\tilde\eps_i)^{-1}\quad&\to\quad \chi_i(t)=\exp(-i\tilde\eps_it),\\
		\tilde G_{k,i}(z)\equiv\tilde\chi_i(z)\frac{iG_{k,i}}{z+i\eps_{k,B}}\quad&\to\quad
		\frac{G_{k,i}}{\tilde\eps_i-\eps_{k,B}}\left( e^{-i\eps_{k,B}t}-e^{-i\tilde\eps_it} \right),\\
		\tilde J_{k,i}(z)\equiv\tilde\chi_i(z)\frac{iJ_{k,i}}{z+i\eps_{k,i}}\quad&\to\quad
		\frac{J_{k,i}}{\tilde\eps_i-\eps_{k,i}}\left( e^{-i\eps_{k,i}t}-e^{-i\tilde\eps_it} \right),\\
		\tilde\chi_{12}(z)\equiv\frac{-\sqrt{\Gamma_{1,B}\Gamma_{2,B}}}{(z+i\tilde\eps_1)(z+i\tilde\eps_2)}
		\quad&\to\quad \chi_{12}(t)
		=\frac{\sqrt{\Gamma_{1,B}\Gamma_{2,B}}}{i(\tilde\eps_1-\tilde\eps_2)}\left( e^{-i\tilde\eps_1t}-e^{-i\tilde\eps_2t} \right),\\
		\tilde \alpha_{k,i}(z)\equiv\tilde\chi_{12}(z)\frac{iG_{k,i}}{z+i\eps_{k,B}}\quad&\to\quad
		-iG_{k,i}\sqrt{\Gamma_{1,B}\Gamma_{2,B}}
		\frac{(\tilde\eps_1-\eps_{k,B})e^{-i\tilde\eps_2t}+(\eps_{k,B}-\tilde\eps_2)e^{-i\tilde\eps_1t}+(\tilde\eps_2-\tilde\eps_1)e^{-i\eps_{k,B}t}}
		{(\tilde\eps_1-\eps_{k,B})(\eps_{k,B}-\tilde\eps_2)(\tilde\eps_2-\tilde\eps_1)}	,\\
		\tilde \beta_{k,i}(z)\equiv\tilde\chi_{12}(z)\frac{iJ_{k,i}}{z+i\eps_{k,i}}\quad&\to\quad
		-iJ_{k,i}\sqrt{\Gamma_{1,B}\Gamma_{2,B}}
		\frac{(\tilde\eps_1-\eps_{k,i})e^{-i\tilde\eps_2t}+(\eps_{k,i}-\tilde\eps_2)e^{-i\tilde\eps_1t}+(\tilde\eps_2-\tilde\eps_1)e^{-i\eps_{k,i}t}}
		{(\tilde\eps_1-\eps_{k,i})(\eps_{k,i}-\tilde\eps_2)(\tilde\eps_2-\tilde\eps_1)}.
	\end{align}
\end{subequations}

\subsection{Inter-dot current}
Let us first evaluate the expectation value of the current operator from site 1 to site 2,
$\hat{I}_{12}=-\sqrt{\Gamma_{1,B}\Gamma_{2,B}}(\hat c_1\dagg \hat c_2+\hat c_2\dagg \hat c_1)/2$
\begin{equation}
	\begin{aligned}
		\langle \hat{\I}_{12}\rangle=-\sqrt{\Gamma_{1,B}\Gamma_{2,B}}
		\Re&\left\{\chi_1^*(t)\chi_{12}(t)\langle \hat c_1\dagg(0)\hat c_1(0)\rangle +\sum_kG_{k,1}^*(t)\left[G_{k,2}(t)+\alpha_{k,1}(t)\right]\langle \hat b_{k,B}\dagg(0)\hat b_{k,B}(0)\rangle\right.\\
		&+\left.\sum_kJ_{k,1}^*(t)\beta_{k,1}(t)\langle \hat b_{k,1}\dagg(0)\hat b_{k,1}(0)\rangle\right\}.
	\end{aligned}
\end{equation}
Like in the main text, we distinguish the exact result from the QME solution by using a calligraphic $\I$.
At late times, only a few terms remain
\begin{equation}
	\langle\hat{\I}_{12}\rangle
	=\sum_k\Re\left\{\Gamma_{1,B}\Gamma_{2,B}
	\left[ \frac{iG_{k,1}^2f_B(\eps_{k,B})}{|\tilde\eps_1-\eps_{k,B}|^2(\eps_{k,B}-\tilde\eps_2)}+\frac{iJ_{k,1}^2f_1(\eps_{k,1})}{|\tilde\eps_1-\eps_{k,1}|^2(\eps_{k,1}-\tilde\eps_2)}\right]
	-\frac{\sqrt{\Gamma_{1,B}\Gamma_{2,B}}G_{k,1}G_{k,2}f_B(\eps_{k,B})}{(\tilde\eps_1^*-\eps_{k,B})(\tilde\eps_2-\eps_{k,B})}\right\}.
\end{equation}
We turn the sum into an integral, noting that we are in the wide-band limit for the reservoir, such that
\begin{equation}
	\begin{aligned}
		\langle\hat{\I}_{12}\rangle&=\Gamma_{1,B}\Gamma_{2,B}\int\frac{d\omega}{2\pi}
		\Re\left\{ \frac{i\Gamma_{1,B}f_B(\omega)}{|\omega-\tilde\eps_1|^2(\omega-\tilde\eps_2)}
		+\frac{i\Gamma_1f_1(\omega)}{|\omega-\tilde\eps_1|^2(\omega-\tilde\eps_2)}
		-\frac{f_B(\omega)}{(\tilde\eps_1^*-\omega)(\tilde\eps_2-\omega)}\right\}\\
		&=\Gamma_{1,B}\Gamma_{2,B}\int\frac{d\omega}{2\pi}
		\left\{ \frac{(\Gamma_2+\Gamma_{2,B})\Gamma_{1,B}f_B(\omega)}{2|\omega-\tilde\eps_1|^2|\omega-\tilde\eps_2|^2}+
		\frac{(\Gamma_2+\Gamma_{2,B})\Gamma_1f_1(\omega)}{2|\omega-\tilde\eps_1|^2|\omega-\tilde\eps_2|^2}
		-\frac{f_B(\omega)}{(\tilde\eps_1^*-\omega)(\tilde\eps_2-\omega)}\right\}
	\end{aligned}
	\label{eq:wide-band_current_integral}
\end{equation}
This current has three parts. The first describes fermions from the joint reservoir entering the double dot on the first site
and being transported to the second site, and the second part is due to electrons entering the system from the first lead (connected to the first site).
Finally, the third term reduces the current $\langle \hat{\I}_{12}\rangle $ and can even make it negative. It arises as a result of fermions added to both sites through the shared reservoir. Their amplitudes add destructively on the second site, but constructively on the first site.
The first and third terms can be made small if the chemical potential of the shared reservoir is lowered.
The second term encodes the desired part of the current.
All parts are also present in \cref{eq:full_QME_current}, where they are encoded as $\gamma_1^+$ and $\gamma_B^+$, which are the rate of electrons being added from the first lead and from the joint reservoir, respectively. 

The integral in \cref{eq:wide-band_current_integral} can be performed numerically for $T\neq0$ and analytically for generic values of the parameters at $T=0$, but the result is cumbersome.
Assuming $\Gamma_i=\Gamma_{\text{lead}}$, $\Gamma_{i,B}=\Gamma_B$, $\delta=\eps_2-\eps_1=0$ and setting temperature $T=0$, we find 
\begin{equation}
	\begin{aligned}
		\langle \hat{\I}_{12}\rangle=
		I_0\left\{ \frac{\Gamma_{\text{lead}}\I_d(V_1)+\Gamma_B\I_d(V_B)}{\Gamma_B+\Gamma_{\text{lead}}}-\I_s(V_B) \right\},
	\end{aligned}
	\label{eq:app_I12}
\end{equation}
where $V_\alpha\equiv2(\mu_\alpha-\eps)/(\Gamma_{\text{lead}}+\Gamma_B)$, 
$I_0=(\Gamma_B\Gamma_{\text{lead}})/(\Gamma_B+\Gamma_{\text{lead}})$, and we define the currents through a single ($s$) and double ($d$) dot (shown in \cref{fig:I12_characteristic})
\begin{equation}
  	\I_s(V)=\frac{1}{2}+\frac{\tan^{-1}(V)}{\pi},\qquad
  	\I_d(V)=\I_s(V)+\frac{V}{\pi(1+V^2)}.
	\label{eq:app_I_defn}
\end{equation}
\begin{figure}[ht]
	\centering
	\includegraphics[width=.4\linewidth]{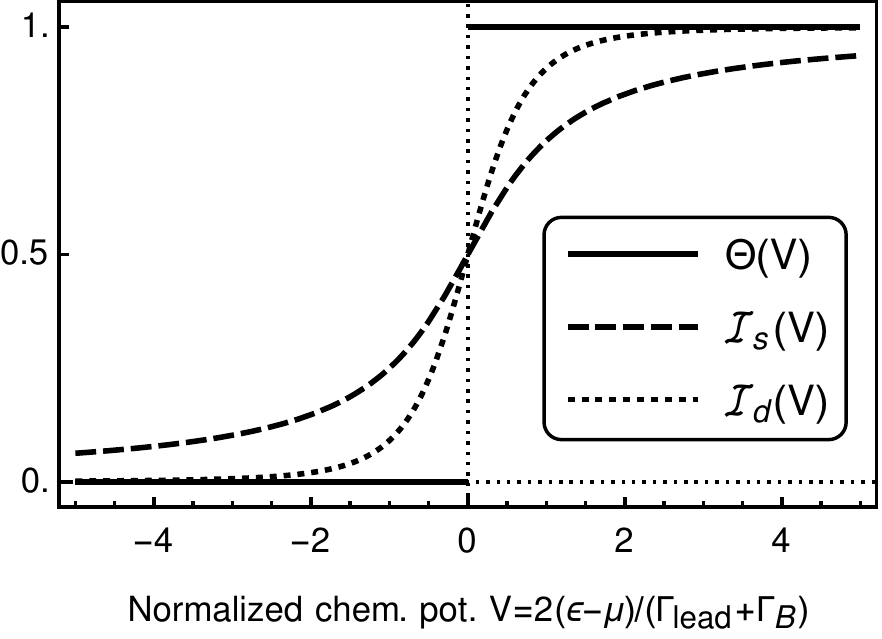}
	\caption{A comparison of the three underlying functions in the current characteristics: $\Theta(V)$ (Heaviside step function), $\I_s(V)$, and $\I_d(V)$ as defined in \cref{eq:app_I_defn}.
	As is discussed in the main text, the Heaviside step function $\Theta(V)$ appears in the QME solution, which does not take the finite width of the modes into account, whereas $\I_s(V)$ and $\I_d(V)$ can be identified as the current through a single mode and two modes.
}
\label{fig:I12_characteristic}
\end{figure}
Taking the chemical potential for the shared reservoir $\mu_{B}\to-\infty$, we are left with the first term, as in the main text.
Finally, here and below, it can be checked that the QME result (at $\delta=0$) can be obtained by replacing
\begin{equation}
	\I_{s,d}(V)\,\rightarrow\, f(\mu-\eps),
	\label{eq:app_replacement}
\end{equation}
which also works for finite $\mu_B$.

\subsection{Current leaving lead 1}
To find the current leaving leads 1 and 2, we consider~\cite{Schaller2014}
\begin{equation}
	\langle\hat{\I}_{i}\rangle=-\lim_{t\to\infty}\frac{d}{dt}\sum_k\langle \hat b_{k,i}\dagg(t) \hat b_{k,i}(t)\rangle.
	\label{eq:lead_current}
\end{equation}
Given the Laplace transform of the system operators \cref{eq:laplace_solution}, we can find the Laplace transform of the reservoir operators \cref{eq:reservoir_modes_laplace}.
Keeping only terms that survive at late times, we obtain
\begin{equation}
	\begin{aligned}
		\hat b_{k,1}(t)\to e^{-i\eps_{k,1}t}\hat b_{k,1}(0)+\frac{J_{k,1}e^{-i\eps_{k,1}t}}{\tilde \eps_1-\eps_{k,1}}\hat c_1(0)
		&+\sum_q\frac{J_{k,1}G_{q,1}}{\eps_{k,1}-\eps_{q,B}}\left( \frac{e^{-i\eps_{q,B}t}}{\tilde\eps_1-\eps_{q,B}}+
		\frac{e^{-i\eps_{k,1}t}}{\eps_{k,1}-\tilde\eps_1}\right)\hat b_{q,B}(0)\\
		&+\sum_q\frac{J_{k,1}J_{q,1}}{\eps_{k,1}-\eps_{q,1}}\left( \frac{e^{-i\eps_{q,1}t}}{\tilde\eps_1-\eps_{q,1}}
		+\frac{e^{-i\eps_{k,1}t}}{\eps_{k,1}-\tilde\eps_1}\right)\hat b_{q,1}(0),
	\end{aligned}
\end{equation}
which gives rise to the reservoir occupation at late times
\begin{equation}
	\begin{aligned}
		\sum_k\langle \hat b_{k,1}\dagg(t)\hat b_{k,1}(t)\rangle &\to\sum_kf_1(\eps_{k,1})
		+\sum_k\frac{J_{k,1}^2}{|\eps_{k,1}-\tilde\eps_1|^2}\langle \hat c_1\dagg(0)\hat c_1(0)\rangle\\
		&-2\Re\sum_{k,q}\frac{J_{k,1}J_{q,1}}{\eps_{k,1}-\eps_{q,1}}e^{i\eps_{k,1}t}\left( \frac{e^{-i\eps_{q,1}t}}{\eps_{q,1}-\tilde\eps_1}-\frac{e^{-i\eps_{k,1}t}}{\eps_{k,1}-\tilde\eps_1} \right)\langle \hat b_{k,1}\dagg(0)\hat b_{q,1}(0)\rangle\\
		&+\int\frac{d\omega\,d\omega'}{4\pi^2}\frac{\Gamma_1\Gamma_{1,B}}{(\omega-\omega')^2}\left|\frac{e^{-i\omega't}}{\omega'-\tilde\eps_1}-\frac{e^{-i\omega t}}{\omega-\tilde\eps_1}\right|^2f_B(\omega')\\
		&+\int\frac{d\omega\,d\omega'}{4\pi^2}\frac{\Gamma_1^2}{(\omega-\omega')^2}\left|\frac{e^{-i\omega't}}{\omega'-\tilde\eps_1}-\frac{e^{-i\omega t}}{\omega-\tilde\eps_1}\right|^2f_1(\omega').
	\end{aligned}
	\label{eq:reservoir_1_occupation}
\end{equation}
The first row is time-independent, so it does not contribute to the current. 
In the second row, both the nominator and denominator go to zero as $q\to k$. Applying l'H\^opital's rule, we find a term linear in $t$, leading to a constant current. 
\begin{equation}
	-2\Re\sum_{k,q}\frac{J_{k,1}J_{q,1}}{\eps_{k,1}-\eps_{q,1}}e^{i\eps_{k,1}t}\left( \frac{e^{-i\eps_{q,1}t}}{\eps_{q,1}-\tilde\eps_1}-\frac{e^{-i\eps_{k,1}t}}{\eps_{k,1}-\tilde\eps_1} \right)\langle \hat b_{k,1}\dagg(0)\hat b_{q,1}(0)\rangle
	\to-2\Re\int\frac{d\omega}{2\pi}\frac{\Gamma_1f_1(\omega)}{(\omega-\tilde\eps_1)^2}\left[ 1+it(\omega-\tilde\eps_1) \right].
\end{equation}
Finally, for the last two rows we need to use \cite{Schaller2014}
\begin{equation}
	\begin{aligned}
		\lim_{t\to\infty}\frac{d}{dt}\int\frac{d\omega}{2\pi}\frac{1}{(\omega-\omega')^2}\left|\frac{e^{-i\omega't}}{\omega'-\tilde\eps_1}-\frac{e^{-i\omega t}}{\omega-\tilde\eps_1}\right|^2
		=\frac{1}{|\omega'-\tilde\eps_1|^2},
	\end{aligned}
\end{equation}
which can be derived from $\lim_{t\to\infty}f(t)=\lim_{z\to0}z\int_0^\infty dt\, e^{-zt}f(t)$~\cite{Schaller2014}.

Hence the current at late times is given by
\begin{equation}
	\langle\hat{\I}_{1}\rangle \to \int\frac{d\omega}{2\pi}\frac{\Gamma_1(\Gamma_1+\Gamma_{1,B})f_1(\omega)}{|\omega-\tilde\eps_1|^2}
	-\int\frac{d\omega}{2\pi}\frac{\Gamma_1\left[ \Gamma_1f_1(\omega)+\Gamma_{1,B}f_B(\omega) \right]}{|\omega-\tilde\eps_1|^2}
	=\int\frac{d\omega}{2\pi}\frac{\Gamma_1\Gamma_{1,B}[f_1(\omega)-f_B(\omega)]}{|\omega-\tilde\eps_1|^2}.
	\label{eq:lead_1_current}
\end{equation}
Again, any reference to lead 2 is absent, because of isolation.
In fact, the form of \cref{eq:lead_1_current} exactly coincides with the current through a single quantum dot connected to two leads, which in this case are the first lead and the shared reservoir.

At zero temperature, we can evaluate the integral straightforwardly to yield
\begin{equation}
	\langle\hat{\I}_{1}\rangle
	=\frac{\Gamma_{1,B}\Gamma_{1}}{\Gamma_{1,B}+\Gamma_1}\left[ \I_s(V_1)-\I_s(V_B) \right]
	\xrightarrow{ \mu_B\to-\infty}\frac{\Gamma_{1,B}\Gamma_1}{\Gamma_{1,B}+\Gamma_1}\I_s(V_1).
	\label{eq:Ilead1}
\end{equation}
$\I_s(V)$ is defined as in the main text [also cf.\ \cref{eq:app_I_defn}]. For a plot see \cref{fig:I12_characteristic}.
Note that the last expression is always positive, so there is no reverse current in the limit $\mu_B\to-\infty$, independent of $\mu_2$.

\subsection{Current leaving lead 2}
We repeat this procedure for the second lead.
We have 
\begin{equation}
	\begin{aligned}
		\tilde b_{k,2}(z)&=\frac{\hat b_{k,2}(0)}{z+i\eps_{k,2}}+\frac{iJ_{k,2}}{(z+i\tilde\eps_2)(z+i\eps_{k,2})}
		\left[ \hat c_2(0)+\sum_q\frac{iG_{q,2}\hat b_{q,B}(0)}{z+i\eps_{q,B}}+\sum_q\frac{iJ_{q,2}\hat b_{q,2}(0)}{z+i\eps_{q,2}} \right]\\
		&-\frac{iJ_{k,2}\sqrt{\Gamma_{1,B}\Gamma_{2,B}}}{(z+i\tilde\eps_2)(z+i\tilde\eps_1)(z+i\eps_{k,2})}
		\left[ \hat c_1(0)+\sum_q\frac{iG_{q,1}\hat b_{q,B}(0)}{z+i\eps_{q,B}}+\sum_q\frac{iJ_{q,1}\hat b_{q,1}(0)}{z+i\eps_{q,1}} \right].
	\end{aligned}
\end{equation}
At late times, this is
\begin{equation}
	\begin{aligned}
		\hat b_{k,2}(t)&\to e^{-i\eps_{k,2}t}\hat b_{k,2}(0)+\frac{J_{k,2}e^{-i\eps_{k,2}t}}{\tilde \eps_2-\eps_{k,2}}\hat c_2(0)
		+\sum_q\frac{J_{k,2}G_{q,2}\hat b_{q,B}(0)}{\eps_{k,2}-\eps_{q,B}}\left( \frac{e^{-i\eps_{q,B}t}}{\tilde\eps_2-\eps_{q,B}}+
		\frac{e^{-i\eps_{k,2}t}}{\eps_{k,2}-\tilde\eps_2}\right)\\
		&+\sum_q\frac{J_{k,2}J_{q,2}\hat b_{q,2}(0)}{\eps_{k,2}-\eps_{q,2}}
		\left( \frac{e^{-i\eps_{q,2}t}}{\tilde\eps_2-\eps_{q,2}}+\frac{e^{-i\eps_{k,2}t}}{\eps_{k,2}-\tilde\eps_2}\right)
		+iJ_{k,2}\sqrt{\Gamma_{1,B}\Gamma_{2,B}}
		\left\{ \frac{\hat c_1(0)e^{-i\eps_{k,2}t}}{(\tilde\eps_1-\eps_{k,2})(\tilde\eps_2-\eps_{k,2})}\right.\\
		&+\left.\sum_q\frac{G_{q,1}\hat b_{q,B}(0)}{\eps_{k,2}-\eps_{q,B}}
		\left[ \frac{e^{-i\eps_{q,B}t}}{(\tilde\eps_1-\eps_{q,B})(\tilde\eps_2-\eps_{q,B})}
		-\frac{e^{-i\eps_{k,2}t}}{(\tilde\eps_1-\eps_{k,2})(\tilde\eps_2-\eps_{k,2})}\right]\right.\\
		&+\left.\sum_q\frac{J_{q,1}\hat b_{q,1}(0)}{\eps_{k,2}-\eps_{q,1}}
		\left[ \frac{e^{-i\eps_{q,1}t}}{(\tilde\eps_1-\eps_{q,1})(\tilde\eps_2-\eps_{q,1})}
		-\frac{e^{-i\eps_{k,2}t}}{(\tilde\eps_1-\eps_{k,2})(\tilde\eps_2-\eps_{k,2})}\right]\right\}.
	\end{aligned}
	\label{eq:reservoir_2_real_time}
\end{equation}
The first four terms are the same as for lead 1, except with $1\leftrightarrow2$. The rest of the expression originates from coupling to site 1.
The reservoir occupation at late times contains the same terms as \cref{eq:reservoir_1_occupation} (except with $1\leftrightarrow2$), in addition to the terms
\begin{equation}
	\begin{aligned}
		&2\Re\int\frac{d\omega\,d\omega'}{4\pi^2}\frac{i\Gamma_{1,B}\Gamma_{2,B}\Gamma_2f_B(\omega')}{(\omega-\omega')^2}
		\left( \frac{e^{i\omega't}}{\tilde\eps_2^*-\omega'}-\frac{e^{i\omega t}}{\tilde\eps_2^*-\omega} \right)
		\left( \frac{e^{-i\omega't}}{(\tilde\eps_1-\omega')(\tilde\eps_2-\omega')}-\frac{e^{-i\omega t}}{(\tilde\eps_1-\omega)(\tilde\eps_2-\omega)} \right)\\
		&+\int\frac{d\omega\,d\omega'}{4\pi^2}\frac{\Gamma_{1,B}^2\Gamma_{2,B}\Gamma_2f_B(\omega')}{(\omega-\omega')^2}
		\left|\frac{e^{-i\omega't}}{(\tilde\eps_1-\omega')(\tilde\eps_2-\omega')}-\frac{e^{-i\omega t}}{(\tilde\eps_1-\omega)(\tilde\eps_2-\omega)}\right|^2\\
		&+\int\frac{d\omega\,d\omega'}{4\pi^2}\frac{\Gamma_1\Gamma_2\Gamma_{1,B}\Gamma_{2,B}f_1(\omega')}{(\omega-\omega')^2}
		\left|\frac{e^{-i\omega't}}{(\tilde\eps_1-\omega')(\tilde\eps_2-\omega')}-\frac{e^{-i\omega t}}{(\tilde\eps_1-\omega)(\tilde\eps_2-\omega)}\right|^2.
	\end{aligned}
	\label{eq:Ilead2_extra}
\end{equation}
The first line originates from the correlator of line 1 and 3 in \cref{eq:reservoir_2_real_time}, whereas the latter two lines stem from the last two lines in \cref{eq:reservoir_2_real_time}.
The time derivative of the first line can be shown to be 
\begin{equation}
	-\frac{d}{dt}\int\frac{d\omega\, d\omega'}{4\pi^2}\Gamma_{1,B}\Gamma_{2,B}\Gamma_2f_B(\omega')\frac{\Gamma_1+\Gamma_{1,B}}{(\omega-\omega')^2}
	\left|\frac{e^{-i\omega't}}{(\tilde\eps_1-\omega')(\tilde\eps_2-\omega')}-\frac{e^{-i\omega t}}{(\tilde\eps_1-\omega)(\tilde\eps_2-\omega)}\right|^2.
\end{equation}
Similarly to above,
\begin{equation}
  	\lim_{t\to\infty}\frac{d}{dt}\int\frac{d\omega}{2\pi}\frac{1}{(\omega-\omega')^2}
  	\left| \frac{e^{-i\omega't}}{(\omega'-\tilde\eps_1)(\omega'-\tilde\eps_2)}-\frac{e^{-i\omega t}}{(\omega-\tilde\eps_1)(\omega-\tilde\eps_2)} \right|^2
  	=\frac{1}{|\omega'-\tilde\eps_1|^2|\omega'-\tilde\eps_2|^2}.
  	\label{}
\end{equation}
Applying the same method as above we derive the current at late times
\begin{equation}
	\langle\hat{\I}_{2}\rangle
	=-\int\frac{d\omega}{2\pi}\frac{\Gamma_2\Gamma_{2,B}[f_B(\omega)-f_2(\omega)]}{|\omega-\tilde\eps_2|^2}
	-\int\frac{d\omega}{2\pi}\frac{\Gamma_{1,B}\Gamma_{2,B}\Gamma_1\Gamma_2}{|\omega-\tilde\eps_1|^2|\omega-\tilde\eps_2|^2}
	\left[f_1(\omega)-f_B(\omega)\right],
\end{equation}
where the first term is the same as for the first lead, except with $1\leftrightarrow2$, whereas the second term is an additional contribution due to the coupling to lead 1.
In the limit of zero-temperature reservoirs, we perform the integral (again setting $\Gamma_{i,B}=\Gamma_B$, $\Gamma_i=\Gamma_{\text{lead}}$ and $\delta=0$)
\begin{equation}
	\langle\hat{\I}_{2}\rangle
	=I_0\left[\I_s(V_2)-\I_s(V_B) \right]
	+\frac{2I_0\Gamma_B\Gamma_{\text{lead}}}{(\Gamma_B+\Gamma_{\text{lead}})^2}\left[ \I_d(V_B)-\I_d(V_1) \right].
\end{equation}

\subsection{Comparison between QME and exact result}
For reference, we collect the expressions for all currents here.
\begin{subequations}
	\begin{align}
  		\langle\hat{\I}_{1}\rangle&=I_0[\I_s(V_1)-\I_s(V_B)],\\
		\langle \hat{\I}_{12}\rangle&=
		I_0\left[ \frac{\Gamma_{\text{lead}}\I_d(V_1)+\Gamma_B\I_d(V_B)}{\Gamma_B+\Gamma_{\text{lead}}}-\I_s(V_B) \right],\\
  		\langle\hat{\I}_{2}\rangle&=
  		I_0\left\{\I_s(V_2)-\I_s(V_B)+\frac{2\Gamma_B\Gamma_{\text{lead}}[\I_d(V_B)-\I_d(V_1)]}{(\Gamma_B+\Gamma_{\text{lead}})^2}\right\}.
	\end{align}
  	\label{eq:app_strong_I}
\end{subequations}
We can compare this with the currents in the weak-coupling limit for the same parameters
\begin{subequations}
	\begin{align}
		\langle \hat I_{1}\rangle_{\text{weak}}&=I_0\left[ f(\eps-\mu_1)-f(\eps-\mu_B) \right],\\
		\langle \hat I_{12}\rangle_{\text{weak}}&=\frac{\Gamma_{\text{lead}}\Gamma_B^2}{(\Gamma_{\text{lead}}+\Gamma_B)^2+\delta^2}\left[f(\eps-\mu_1)-f(\eps-\mu_B)\right],\\
		\langle \hat I_{2}\rangle_{\text{weak}}
		&=I_0\left[ f(\eps-\mu_2)-f(\eps-\mu_B) \right]
		+\frac{2I_0\Gamma_B\Gamma_{\text{lead}}}{(\Gamma_{\text{lead}}+\Gamma_B)^2+\delta^2}\left[ f(\eps-\mu_B)-f(\eps-\mu_1) \right].
	\end{align}
\end{subequations}
It is straightforward to verify the replacement in the main text even for finite $\mu_B$ (but still $\delta=0$).

The current leaving lead 1 that does not enter lead 2 flows into the shared reservoir
$\langle\hat{\I}_{\text{B}}\rangle=-\langle\hat{\I}_{1}\rangle-\langle\hat{\I}_{2}\rangle$.

\end{widetext}

\bibliographystyle{apsrev4-1}
\bibliography{library}{}
\end{document}